\documentclass[aps,prb,twocolumn]{revtex4}

\usepackage{amsmath}
\usepackage{graphicx}
\usepackage{bm}
\usepackage{color}

\begin{document}

\title{Lyapunov exponents of heavy particles in turbulence}

\author{J\'er\'emie Bec} \affiliation{CNRS UMR6202, Observatoire de la C\^ote
d'Azur, B.P.\ 4229, 06304 Nice Cedex 4, France.}

\author{Luca Biferale} \affiliation{Dept. of Physics and INFN,
University of Rome ``Tor Vergata'', Via della Ricerca Scientifica 1,
00133 Roma, Italy.}

\author{Guido Boffetta}
\affiliation{Dip. di Fisica Generale and INFN,
University of Torino, Via Pietro Giuria 1, 10125, Torino, Italy, \\
and CNR-ISAC c.Fiume 4, 10133 Torino Italy.}

\author{Massimo Cencini} \affiliation{INFM-CNR, SMC Dipartimento di
  Fisica, Universit\`a di Roma ``La Sapienza'', Piazzale A.\ Moro 2,
  I-00185 Roma, Italy and ISC-CNR Via dei Taurini 19, I-00185 Roma,
  Italy.}

\author{Stefano Musacchio}
\affiliation{CNRS, INLN, 1361 Route des Lucioles, F-06560 Valbonne, France.}

\author{Federico Toschi} \affiliation{CNR-IAC, Viale del Policlinico 137,
  I-00161 Roma, Italy, and INFN, Sezione di Ferrara, via G. Saragat 1,
  I-44100, Ferrara, Italy.}

\date{\today}
\begin{abstract}
Lyapunov exponents of heavy particles and tracers advected by
homogeneous and isotropic turbulent flows are investigated by means of
direct numerical simulations.  For large values of the Stokes number,
the main effect of inertia is to reduce the chaoticity with respect to
fluid tracers. Conversely, for small inertia, a counter-intuitive
increase of the first Lyapunov exponent is observed. The flow
intermittency is found to induce a Reynolds number dependency for the
statistics of the finite time Lyapunov exponents of tracers. Such
intermittency effects are found to persist at increasing inertia.
\end{abstract}
\maketitle

Impurities suspended in incompressible flows are relevant to several
physical processes, such as spray combustion, raindrop formation and
transport of pollutants.\cite{PA02,FFS02,Seinfeld} These particles,
being typically of finite size and heavier than the ambient fluid, cannot
be modeled as simple tracers. They indeed posses inertia, which is
responsible for the spontaneous generation of strong inhomogeneities
in their spatial distribution.\cite{EF94,EKR96,FP04} In a turbulent flow, their
clustering is more efficient at small scales, below the dissipative
scale, where the fluid velocity field is smooth. The intensity
of particle clustering is there related to the statistics of the
Lyapunov exponents of particle trajectories.\cite{B03}
The behavior of the Lyapunov exponents of
inertial particles and the relation with particle clustering was
recently investigated in random short-correlated
flows.\cite{WM05,DMOW05,BCH06}

In this Letter, be means of high-resolution direct numerical simulations, 
we investigate the Lyapunov spectra of inertial particles, 
varying their response time $\tau_s$ and the Reynolds number of the
carrier turbulent flow.  For $\tau_s$ larger than the Kolmogorov-scale
turnover time $\tau_\eta$, the presence of inertia results in a
generic reduction of chaoticity: the leading Lyapunov Exponent is
smaller than the one of tracers and the strongest chaotic fluctuations
become less probable.  Remarkably, for $\tau_s<\tau_\eta$, an
increase of chaoticity is observed; this effect can be understood in
terms of the preferential concentration of particles in the
high-strain regions of the flow.  
For fluid tracers, we observe
intermittency effects on the statistics of chaotic fluctuations: the
Reynolds-number dependence deviates from the dimensional prediction.
Such deviations are found to persist for particles with inertia.


A small spherical particle of radius $a$ with a density
$\rho$ much larger than the density $\rho_0$ of the surrounding incompressible
fluid, evolves according to the dynamics\cite{MR83}
\begin{equation}
\dot{\bm x} = {\bm v}\, ,\quad \dot{\bm v} = - \frac{1}{\tau_{s}}
\left[{\bm v} - {\bm u}({\bm x(t)},t) \right]\, ,
\label{eq:1}
\end{equation}
where ${\bm u}$ is the fluid velocity at the location $\bm x$ of the
particle that moves with velocity ${\bm v}$, and $\tau_{s}\!=\!2 a^2
\rho/(9 \nu \rho_0)$ is the Stokes time ($\nu$ is the kinematic
viscosity of the fluid).  Equation~(\ref{eq:1}) holds when the flow
surrounding the particle is a Stokes flow; this requires $a \ll \eta$,
$\eta$ being the Kolmogorov scale of the flow. The Stokes number is
defined as $St=\tau_s/\tau_\eta$, where $\tau_\eta$ is the eddy
turnover time associated to $\eta$.  Particles are assumed to behave
passively: we neglect their feedback on the flow, which is justified
for very diluted suspensions.\cite{FKE94} For comparison, we also
study the motion of neutral particles that follow the dynamics
$\dot{\bm x}={\bm u}({\bm x(t)},t)$, which corresponds to the limit
$\tau_{s} \to 0$ in Eq.~(\ref{eq:1}).

The incompressible fluid velocity field ${\bm u}({\bm x},t)$ evolves
according to the Navier--Stokes equations
\begin{equation}
{\partial}_t {\bm u} + {\bm u} \cdot {\bm \nabla} {\bm u} = -
   (1/\rho_0)\, {\bm \nabla} p + \nu \Delta {\bm u} + {\bm f}\, ,
   \quad {\bm \nabla}\cdot {\bm u} = 0\, ,
\label{eq:2}
\end{equation}
where the forcing provides an external energy input at a rate
$\varepsilon=\langle {\bm f}\cdot{\bm u} \rangle$.  These equations
are integrated on $d=3$ dimensional grid of size $N=128,\,256,\,512$
with periodic boundary conditions, by means of a fully de-aliased
pseudo-spectral parallel code with 2$^{\rm nd}$ order Adams--Bashforth
time-stepping.  Energy is injected by keeping constant the spectral
content of the two smallest wavenumber shells. Viscosity is chosen to
resolve well the small-scale velocity ($k_{max} \eta \simeq 1.7$). 
The Reynolds numbers based on Taylor's micro-scale is
in the range $R_{\lambda} \in [65: 185]$ (see
Ref.~\onlinecite{BBBCCLMT06} for further details).

Once the fluid flow has reached a statistically stationary state,
particles are homogeneously seeded with initial velocities equal to
the fluid velocity at their locations. We followed 33 sets of 2000
particles with Stokes numbers ranging from $0$ to $2.2$ for a time
$\simeq 200\,\tau_\eta$ after relaxation of transients.  

In order to compute the Lyapunov spectrum we follow along each particle
trajectory the time evolution of $2\times d$ infinitesimal
displacements in the position-velocity phase space obtained by
linearizing the particle dynamics (\ref{eq:1}).  The infinitesimal
volume $V^j$, defined by $j$ linear independent tangent vectors, grows
in time with an exponential rate $\sum_{i=1}^{j} \gamma_i(T)=
({1}/{T}) \ln \left[ {V^j(T)}/{V^j(0)}\right]$, which defines the
\emph{finite-time Lyapunov exponents} $\gamma_{i}(T)$ (FTLE), also
called stretching rates. They asymptotically converge to the Lyapunov
exponents $\lambda_i=\lim_{T\to \infty} \gamma_i(T)$, which by definition are
labeled in decreasing order $\lambda_1\ge \cdots \ge \lambda_{2d}$.
To compute these exponents numerically, we make use at each time lag
$\tau_{\eta}$ of a standard technique based on the orthonormalization
of the infinitesimal displacements by a Gram--Schmidt procedure (see,
e.g., Ref.~\onlinecite{CPVbook}).

\begin{figure}[h!]
\centerline{
\includegraphics[width=0.5\textwidth,draft=false]{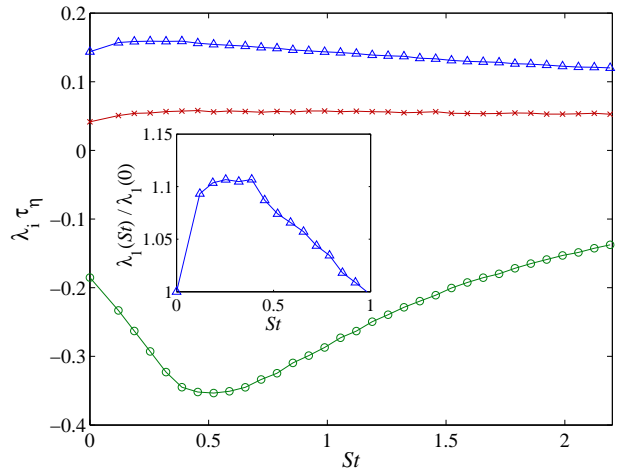}}
\vspace{-10pt}
\caption{Lyapunov exponents $\lambda_i$ for $i=1$ (triangles), $i=2$
  (crosses) and $i=3$ (circles) as a function of Stokes
  number. $R_{\lambda}= 185$.  In the inset we show the relative
  growth of the first Lyapunov exponent $\lambda_1(St)/ \lambda_1(0)$
  occurring at small $St$.  }
\label{lyap3_stokes}
\end{figure}

In Fig.~\ref{lyap3_stokes} we show the behavior of the first three
Lyapunov exponents as a function of $St$ for the largest value of the
Reynolds number.  These three exponents rule the time evolution of
infinitesimal elements in the physical space. For the range of Stokes
numbers investigated here, we observe $\lambda_4\sim \lambda_5\sim
\lambda_6\approx -1/\tau_s$, signaling the relaxation of particle
velocities to the fluid. 

The largest Lyapunov exponent $\lambda_1$ measures the chaotic separation of
particle trajectories. To understand how chaoticity is affected by inertia,
two mechanisms have to be considered. First, inertial particles have a
delay on the fluid motion; this means that their velocity is
approximately given by that of tracers with a time
filtering over a time window of size $\tau_s$. This
effect weakens chaoticity. Second, heavy particles are ejected from
persistent vortical structures and concentrate in high-strain regions.
Since these portions of the flow are characterized by larger
stretching rates, the chaoticity of particle trajectories is increased with
respect to tracers that homogeneously visit all the regions. As
emphasized in the inset of Fig.~\ref{lyap3_stokes} the latter effect
dominates for $St < 1$ where $\lambda_1$ is larger than the Lyapunov
exponent of the fluid tracers ($St = 0$).  This is not be predicted
from analytical and numerical studies done in white-in-time random
velocity fields:\cite{DMOW05,BCH06} such flows clearly possess no
persistent structures.  Conversely, at sufficiently large $St$, the
Lyapunov exponent decreases: preferential concentration is then
negligible and the time-filtering approximation becomes relevant. For
such a large inertia, the white-in-time models apply and indeed
predict a decrease of $\lambda_1$ as a function of $St$.
Note that the competition between filtering and preferential
concentration described above also enters in the distribution of particle
acceleration.\cite{BBBCCLMT06}

\begin{figure}[h!]
\centerline{
\includegraphics[width=0.5\textwidth,draft=false]{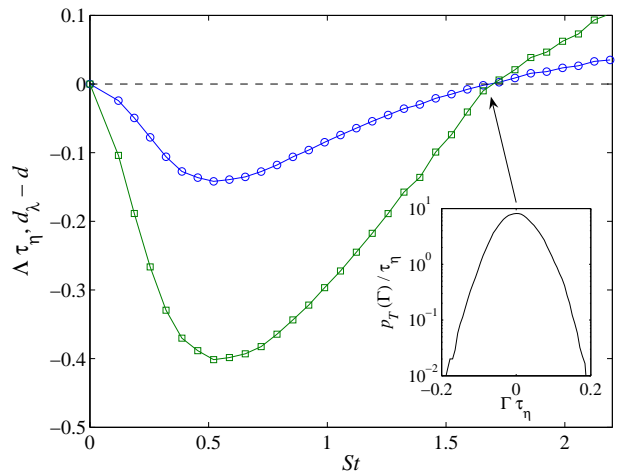}}
\vspace{-10pt}
\caption{Lyapunov dimension $d_{\lambda}$ (squares) and volume growth
rate $\Lambda = \sum_{i=1}^3 \lambda_i$ (circles) as a function of
Stokes number ($R_{\lambda}= 185$).  Inset: PDF of the finite-time
volume growth rate $\Gamma(T)$ for $T\approx 80\,\tau_\eta$.}
\label{dimenlyap}
\end{figure}

As observed in Fig.~\ref{lyap3_stokes}, the time evolution of
infinitesimal surfaces is also affected by these two mechanisms.
Indeed, at varying $St$, the second Lyapunov exponent $\lambda_2$
displays a behavior qualitatively similar to that of $\lambda_1$. For
the tracers ($St=0$), incompressibility of the flow implies
$\lambda_1+\lambda_2+\lambda_3=0$. Since for a time reversible
dynamics one has $\lambda_2=0$, the ratio $\lambda_2/\lambda_1$ is a
measure of the irreversibility of the dynamics.  Previous
numerical investigations at moderate Reynolds numbers\cite{GP90}
indicate $\lambda_2/\lambda_1 \simeq 0.25$; our simulations indicate
$\lambda_2/\lambda_1 = (0.28 \pm 0.02) $.  This irreversibility stems
from the fact that the Navier--Stokes equation itself is not invariant
with respect to time reversal.

\begin{figure}[h!]
  \centerline{
    \includegraphics[width=0.5\textwidth,draft=false]{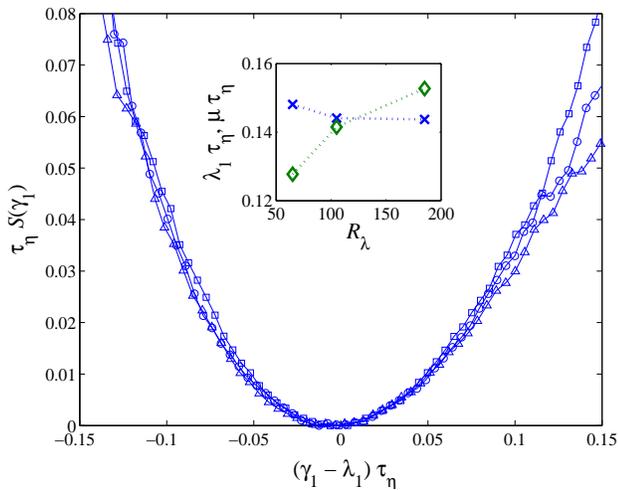}}
  \vspace{-10pt}
  \caption{Cramer function $S(\gamma_1)$ for fluid tracers for for
    $R_{\lambda} = 65$ (squares), $105$ (circles) and $185$
    (triangles). Inset: Lyapunov exponent $\lambda_1$ (crosses) and
    reduced variance $\mu = T\sigma^2$ (diamonds) as a function of
    $R_\lambda$.}
\label{pdfs}
\end{figure}

The behavior of $\lambda_3$ as a function of $St$ markedly differs
from that of the first two exponents. Due to the dissipative nature of
the inertial particle dynamics, volumes in physical space are not
conserved. Indeed, the volume growth rate, defined as $\Lambda =
\lambda_1+\lambda_2+\lambda_3$, which identically vanishes for fluid
tracers, is negative for all Stokes numbers in the range $0<St\lesssim
1.72$ (see Fig.~\ref{dimenlyap}). This means that all volumes from
physical space contract to zero at large times. Such clustering, which
happens at scales much smaller than $\eta$, is a consequence of the
convergence of particle trajectories toward a dynamically evolving
(multi)fractal set.\cite{B03} The fractal dimension of
this attractor can be estimated by means of the Kaplan--Yorke (or
Lyapunov) dimension\cite{ER85} as $d_{\lambda} = J + \sum_{i=1}^{J}
\lambda_i / |\lambda_{J+1}| $ where $J$ is the largest integer such
that $\sum_{i=1}^{J} \lambda_i > 0 $. The fractal dimension of
inertial particles is represented in Fig.~\ref{dimenlyap} as a
function of $St$.  The minimum at $St\approx 0.5$ corresponds to
maximal clustering.  For $St\gtrsim 1.72$ the fractal dimension
becomes greater than $d=3$, indicating that the spatial distribution
of particles is not fractal anymore. For $St\approx 1.72$ the volume
growth rate $\Lambda$ vanishes, meaning that the dynamics of such
particles preserve volumes on average. However, the finite-time volume
growth rate $\Gamma = \gamma_1 + \gamma_2 + \gamma_3$ experiences
large fluctuations, as shown in the inset of Fig.~\ref{dimenlyap}.  As
a result, strong local inhomogeneities are present in the particle
concentration also at large values of $St$.


We now turn to the study of the Lyapunov exponent dependence 
on the Reynolds number of the flow. The inset of
Fig.~\ref{pdfs} shows the first Lyapunov exponent for neutral
particles (i.e.\ $St=0$) as a function of $R_{\lambda}$. 
Since $\lambda_1$ is a small-scale turbulent quantity
with the dimension of an inverse time, one expects $\lambda_1
\tau_{\eta} \simeq const$ and thus $\lambda_1 \propto \ R_{\lambda}$.
However, as a consequence of the intermittent
fluctuations of the velocity gradients in turbulent flows, one can
predict an anomalous dependence on Reynolds number $\lambda_1 \propto
R_{\lambda}^{\alpha}$ with $\alpha<1$ \cite{CJPV93}. 
This implies that $\lambda_1 \tau_{\eta}$
decreases with Reynolds, as indeed confirmed by our simulations.

\begin{figure}[h!]
\centerline{
\includegraphics[width=0.5\textwidth,draft=false]{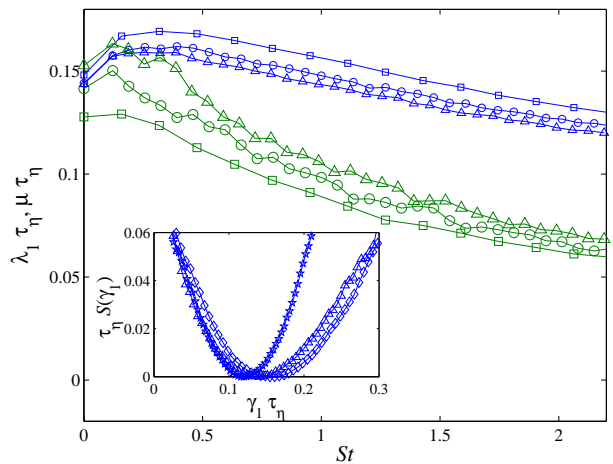}}
\vspace{-10pt}
\caption{Mean value $\lambda_1$ (small symbols) and reduced variance
$\mu$ (large symbols) of the FTLE as a function of the Stokes number,
for $R_{\lambda} = 65$ (squares), $R_{\lambda} = 105$ (circles) and
$R_{\lambda} = 185$ (triangle). Inset: Cramer function $S(\gamma_1)$
for $R_\lambda\approx185$, and various values of the Stokes number:
$St=0$ (triangles), $St=0.32$ (diamonds) and $St=2.19$ (stars).}
\label{lyapmu_stokes}
\end{figure}

Intermittency is actually expected to affect the whole probability
distribution function (PDF) of the largest finite-time Lyapunov
exponents $\gamma_1(T)$.  For $T$ sufficiently large, the distribution
of FTLE is expected to obey a large-deviation principle,
i.e. $p_T(\gamma_1) \propto \exp({-T\,S(\gamma_1)})$. The
Cram\'er (or rate) function $S(\gamma_1)$ is a non-negative concave
function that vanishes at its minimum, attained for
$\gamma_1=\lambda_1$.  
Small fluctuations occurring when $|\gamma_1-\lambda_1| \ll T^{-1/2}$ are
described by the central-limit theorem which amounts to approximating
$S(\cdot)$ by a parabola in the vicinity of its minimum.  The effect
of intermittency on such small fluctuations can be measured from the
variance $\sigma^2 \equiv \langle (\gamma_1- \lambda_1)^2 \rangle$ of
the FTLE, or more particularly from the reduced variance $\mu = T
\sigma^2$ which measures the width of the Cram\'er function.  As
predicted in Ref.~\onlinecite{CJPV93}, intermittency is responsible
for an anomalous dependency of $\mu$ on the Reynolds number.  More
particularly $\mu \tau_{\eta}$ is expected to grow with $R_{\lambda}$.
This tendency is qualitatively confirmed by our simulations as shown
on the inset of Fig.~\ref{pdfs}. The signature of intermittency on the
higher-order statistics of $\gamma_1$ can hardly be measured in a
reliable way. Indeed, as shown in Fig.~\ref{pdfs} the PDFs of
$\gamma_1$ for the three $R_\lambda$ considered, once centered and
normalized, almost collapse for fluctuations as large as $3\sigma$.
However it is still possible to observe a systematic deviation from
the Gaussian distribution. Because of incompressibility, the left tail
of the PDF is bounded by the constraint that $\gamma_1>0$. It thus has
to decrease faster than a Gaussian as indeed observed.  The right part
of the PDF is related to strong velocity gradients.  Such events
apparently lead to a tail which is fatter than Gaussian.  Turbulent
intermittency should therefore mainly affect the right tail.


For inertial particles, intermittent corrections act in the same
direction as for fluid tracers.  Figure \ref{lyapmu_stokes} shows the
behavior of $\lambda_1$ and of the reduced variance $\mu$ as a
function of the Stokes number for various Reynolds numbers. As for
tracers, for any given $St$, $\lambda_1 \tau_{\eta}$ decreases while
$\mu \tau_{\eta}$ increases with $R_{\lambda}$.  The inset of
Fig.~\ref{lyapmu_stokes} shows, for $R_\lambda = 185$, the Cramer
function of the FTLE $\gamma_1(T)$ for both neutral particles ($St=0$)
and inertial particles with two different $St$. For $St<1$, the whole
distribution shifts to higher values, signaling the increased
chaoticity. For $St>1$, the distribution of FTLE shifts to lower
values and fluctuations becomes less probable.  The asymmetry
observed in the PDF of tracer stretching rates decreases with inertia.
At the largest values of $St$ which are considered, the Cramer
function $\gamma_1$ becomes indistinguishable from a parabola.
Finally, it is worth noticing that the dependence on the Reynolds
number is less significant for the volume growth rate $\Lambda$, and
thus for the fractal dimension.  Intermittency affects only
weakly particle clustering.  A clearer understanding of this issue
requires to investigate larger values of the Reynolds number with
longer statistics.  This can hardly be achieved numerically but could
be done experimentally using standard box-counting techniques to
estimate the fractal dimensions of the particle spatial distribution.


We have seen that two mechanisms enter the dynamics of inertial
particles: they concentrate in high-strain regions and they lag behind
the fluid flow. Only in either the limit of small or large inertia,
one of the two effect dominates the other. At present time, tackling
analytically the behavior of the largest Lyapunov exponent as a
function of the Stokes number could only be done in these
asymptotics.\cite{DMOW05} For the range of Stokes numbers
considered here, both effects compete and influence the Lyapunov
exponents, preventing a complete analytical description.
A numerical confirmation of the present theoretical predictions
would require larger computational resources.

We acknowledge useful discussions with A. Celani and A. Lanotte.
This work has been partially supported by the EU under contract
HPRN-CT-2002-00300 and the Galileo program on Lagrangian
Turbulence. Numerical simulations have been performed at CINECA
(Italy) and IDRIS (France) under the HPC-Europa program, contract
number RII3-CT-2003-506079.



\end{document}